\documentclass[aps,pra,amsmath,amssymb,twocolumn]{revtex4}
\usepackage{amssymb}
\usepackage{color}
\usepackage{graphicx}
\usepackage{epsfig,amssymb,amsmath}

\usepackage{color}

\def\>{\rangle}
\def\<{\langle}
\def\comment#1{ [{\bf Comment:} {\sf #1}]}
\def\labell#1{\label{#1}}
\def\togli#1{}

\begin{document}

\title{Quantum radar}
\author{Lorenzo Maccone$^1$ and Changliang Ren$^2$
}\affiliation{\vbox{1.~Dip.~Fisica and INFN Sez.\ Pavia, University
    of Pavia, via Bassi 6, I-27100 Pavia, Italy}\\
  \vbox{2.~Center for Nanofabrication and System Integration,
    Chongqing Institute of Green and} \vbox{ Intelligent Technology,
  Chinese Academy of Sciences, Chongqing 400714, People's
    Republic of China}} 
\begin{abstract}
  We propose a quantum metrology protocol for the localization of a
  non-cooperative point-like target in three-dimensional space, by
  illuminating it with electromagnetic waves.  It employs all the
  spatial degrees of freedom of $N$ entangled photons to achieve an
  uncertainty in localization that is $\sqrt{N}$ times smaller for
  each spatial direction than what could be achieved by $N$
  independent photons.
\end{abstract}
\maketitle

Quantum metrology \cite{review,qmetr,caves,rafalreview,matteo} is a
set of procedures that increase the precision in the estimations of
parameters by employing quantum effects such as entanglement or
squeezing. By entangling $N$ different probes, typical protocols
achieve a $\sqrt{N}$ decrease in the statistical noise over what would
be achievable without entanglement. Here we will present a quantum
metrology protocol for a radar. Radar stands for RAdio Detection And
Ranging, so the bare minimum for a protocol to qualify as such is that
it is able to detect a target and return its position relative to the
receiver. However, previous quantum radar protocols
\cite{qradarqillum}, based on quantum illumination \cite{qillum} fail
this requirement as they can only discriminate whether the target is
present or not, and they give a quantum advantage only in the presence
of a rather specific thermal noise model.  Other protocols
\cite{lanzag,qradarothers} still are unable to provide {\em both}
detection and position of the target with enhanced precision. Here we
will present a quantum metrology protocol for a radar.  Instead our
protocol returns both and does not require the target to cooperate. It
achieves a $N^{3/2}$ decrease in the uncertainty volume of the target
position over what could be achieved with $N$ independent photons of
the same spatial bandwidth, namely a $\sqrt{N}$ decrease in
uncertainty along each of the three spatial dimensions. The main
drawbacks of our protocol are the difficulty in creating the required
entangled state of the electromagnetic field and its sensitivity to
noise. Regarding the first problem, according to current technologies,
we discuss how at least the case of $N=2$ can be experimentally
realized through spontaneous parametric down-conversion under a
tightly focused pulse pump based on type II noncritical phase
matching.  Regarding the second, known techniques
(e.g.~\cite{qps1,changliang}) can be adapted here, leading to a
reduction of the impact of noise with a slight decrease in
performance.

The main idea of our protocol is to combine a three-dimensional
generalization of the one-dimensional quantum localization protocol of
\cite{qps,qps1} with a free-space propagation analysis of the signal
from target to receiver. The use of all the spatial degrees of freedom
of the entangled photons allows three dimensional localization.

The paper's outline follows. To simplify the discussion, we will first
present the case of two photons, and then give the $N$-photon general
protocol. We start with the case of two maximally-entangled photons.
Then we show that, while a reduction of entanglement entails a
reduction of precision, it also decreases the transverse dimensions of
the required detector. We conclude by providing some modifications of
the protocol that strengthen the protocol against the effects of
noise.

\begin{figure}[h!]
\vspace{-.2cm}
\begin{center}
\epsfxsize=.75\hsize\leavevmode\epsffile{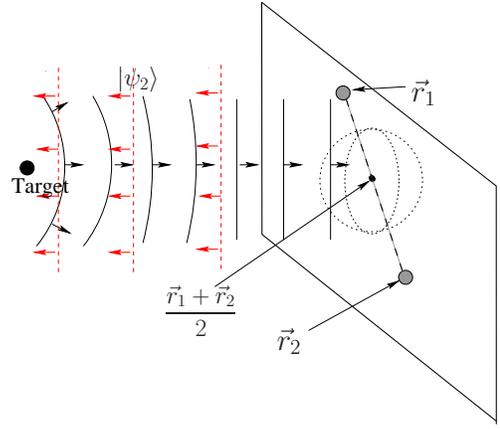}
\end{center}
\vspace{-.5cm}
\caption{Quantum radar setup (two-photon case). A point-like target reflects the
  momentum-entangled state $|\psi_2\>$ of two photons (impinging
  dashed lines). In the far field, the photons arrive at a screen. The
  average time of arrival (not pictured) provides the longitudinal
  distance, whereas the average of the two photons transverse arrival
  positions $\vec r_1$, $\vec r_2$ provides the object transverse
  location (dashed line). The uncertainty sphere obtained (dotted
  line) is reduced by a factor $N^{3/2}$ over what would be obtained
  with $N$ independent photons with same spatiotemporal bandwidth (the
  case $N=2$ is depicted here).
  \label{f:fig}}\end{figure}

The protocol allows a receiver to find her position relative to an
uncooperating target object that is illuminated with a suitable
entangled state of light composed of $N$ entangled photons, see
Fig.~\ref{f:fig}. To this aim the receiver measures their arrival
position and arrival time on a transverse plane at her location.
Consider $N=2$ first. The joint probability of photodetection, namely
of finding the two photons at times ${t}_1$, ${t}_2$ and at positions
$\vec r_1$, $\vec r_2$ (two-dimensional transverse vectors) is
\cite{mandel}
\begin{align}
  p({t}_1,\vec r_1;{t}_2,\vec r_2)\propto|\<0|E^+({t}_1,\vec
  r_1)\:E^+({t}_2,\vec r_2)|\psi_2\>|^2
  \labell{joint}\;,
\end{align}
where $|0\>$ is the vacuum state, $|\psi_2\>$ is the state of the two
photons (we work in the Heisenberg picture where the operators evolve
from an initial time ${t}_0$) and (e.g.~\cite{shih})
\begin{align}
  E^+({t},\vec r)\equiv\int d\vec k_3\;g(\vec k_3,{t},\vec r)\;a(\vec
  k_3)\:e^{i\omega ({t}-{t}_0)}
\labell{epiu}\;,
\end{align}
where $g$ is the transfer function (defined below) between the object
plane (at the target's position) and the image plane (at the position
of the receiver), $a(\vec k_3)$ is the electromagnetic field
annihilation operator for the mode with wave vector $\vec k_3$. As
customary, we will employ the far-field approximation, valid when the
object-receiver distance is sufficiently large. In this case, the
longitudinal component of the wave vector is much larger than the
transverse components: $k_x^2+k_y^2\ll |\vec k_3|^2$, where $|\vec
k_3|=({k_x^2+k_y^2+k_z^2})^{1/2}=\omega/c$, with $\omega$ the light's
frequency. So the $\vec k_3$ integral can be approximated as
\begin{align}
  \int d\vec k_3=\int \frac{d\omega}{c^2}d\vec k/\sqrt{\tfrac
    1{c^2}-\tfrac{k_x^2+k_y^2}{\omega^2}}\simeq \tfrac 1c\int d\omega\,d\vec
  k
\labell{ff}\;,
\end{align}
with $\vec k$ the two dimensional transverse wave vector $\vec
k=(k_x,k_y)$.  Then, Eq.~\eqref{epiu} can be replaced by \begin{align}
  E^+({t},\vec r) \simeq\int d\omega\:d\vec k \;g(\vec k,\vec
  r)\:a(\omega,\vec k)\:e^{i\omega({t}-{t}_0)}
\labell{epl}\;,
\end{align}
where the longitudinal component contributes only with a phase factor
which measures the longitudinal distance $z=c({t}-{t}_0)$ that the
light travels from the source to the target, and back to the detector,
and where the free-space (transverse) transfer function is
\begin{align}
  g(\vec k,\vec r)\equiv\int d\vec r_o\:A(\vec
  r_o)\:e^{i\vec k\cdot(\vec r_0-\vec r)}
\labell{g}\;,
\end{align}
where $A$ is the object transfer function and the integral is over the
(transverse) object plane, namely $\vec r_o$, $\vec r$ are
two-dimensional transverse vectors. We will consider a point-like
reflective object which reflects only the photons that impinge on its
position $\vec r_p$. The other photons are lost.
 This situation is
described\togli{is it true? Does this transfer function describe
  this situation? Does the event of receiving at the detector photons
  that are in the input state without reflection have a negligible
  probability? This is probably INCORRECT!!! because
\begin{align}
|\psi_2\>=\int dt_1\:d\vec r_1\:dt_2\:d\vec r_2\nonumber
\:\tilde\psi({t}_1+{t}_2,\vec r_1+\vec r_2)\:\\\times
a^\dag\!({t}_1,\vec r_1)
\:a^\dag\!({t}_2,\vec r_2)|0\>
\labell{a}\;
\end{align}means that the photons in the state arrive anywhere! We
need some way to discriminate the photons that are reflected off the
target from the ones in the initial state: perhaps using the
polarization degree of freedom?!? We could hypothesize that
polarization is rotated during a reflection? Otherwise we have to drop
the claim that the target is uncooperative and we have to require that
Bob actively rotates the polarization... Otherwise, perhaps we can
claim that the state $|\psi_2\>$ does not arrive at the receiver's location,
but we need to check if this is consistent with the rest of our
claims. YES: we can claim this by saying that the state approximates
$|\psi\>$ at Bob's location: Alice's very distant.
}
by a transfer function which has value $a$ in the vicinity of $\vec
r_p$ in the object plane, and value zero elsewhere in the object
plane, namely $A(\vec r_o)\propto a\,\delta(\vec r_o-\vec r_p)$.
Slightly more general situations can be considered, but it is not
possible to perform more complex imaging with entangled light since
the transfer function $g$ of any imaging apparatus is more complex
than \eqref{g} and the photon correlations in \eqref{psi} (below) will
prevent the formation of a discernible image. For quantum radar
applications, we are only interested in free-space propagation,
described by \eqref{g} and in detection and ranging, rather than
imaging.

The necessary entangled two-photon state, produced at the initial time
${t}_0$, in the far-field approximation, is
\begin{align}
  |\psi_2\>\equiv\int d\omega\:d\vec k\;\psi(\omega,\vec
  k)(a^\dag(\omega,\vec k))^2|0\>
  \labell{psi}\;
\end{align}
where $a^\dag(\omega,\vec k)$ creates a photon with frequency $\omega$
and transverse wave vector $\vec k$, $\psi$ is the biphoton's
spatiotemporal wavefunction and we omit the normalization since it is
a non-normalizable state as all EPR states \cite{epr}. It is a
maximally-entangled state in three different degrees of freedom:
$k_x$, $k_y$ and $\omega$ (we will drop this assumption later). We
must also suppose that at the receiver's location there is a
negligible probability of seeing the photons that are not scattered by
the object, namely \eqref{psi} is an approximation of the
electromagnetic field valid only in the object's vicinity. In practice
this can be implemented by requiring that the longitudinal component
of the wave vector $\vec k_3$ is directed away from the detector
(which is implicit in the far field approximation).

Replacing these quantities into Eq.~\eqref{joint}, we find
\begin{align}
  p({t}_1,\vec r_1;{t}_2,\vec
  r_2)\propto|\tilde\psi({t}_1+{t}_2-2t_0,\vec r_1+\vec r_2-2\vec r_p)|^2
\labell{fin2}\;,
\end{align}
where $\tilde\psi({t},\vec r)=\int d\omega\:d\vec k\:\psi(\omega,\vec
k)\:e^{i(\omega {t}+\vec k\cdot\vec r)}$ is the Fourier transform of
$\psi(\omega,\vec k)$. This implies that the average time of the
arrival is equal to the transit time of the signal from its production
at ${t}_0$ to its detection at $t$,\togli{What is the time ${t}=0$
  physically? What is the time at which I start the clocks at the
  detection? We cannot use the time at which the signal interacts with
  the object as ${t}=0$, because that time in unknown!  No ${t}=0$,
  which I called ${t}_0$ is the initial state when the beam is created
  because we're working in the Heisenberg picture.}  and that the
average arrival transverse position is equal to the object's
transverse position. The statistical noise of these two quantities is
given by {\em half} the standard deviation of $|\tilde\psi|^2$ in time
and in position.  Indeed, the left-hand-side of \eqref{fin2} can also
be written as $ |\tilde\psi\left(2(\tfrac{{t}_1+{t}_2}2-{t}_0),
  2(\tfrac{\vec{r}_1+\vec{r}_2}2-\vec{r}_0)\right)|^2$. Hence, the
standard deviation of the average time of arrival gains a factor of
$1/2$, and similarly for each of the two components of the average
position.

Naturally, we must compare this result to what one can obtain using two unentangled
photons with the same spectral characteristics. Consider then a single
photon in the state
\begin{align}
|\psi_1\>=\int d\omega\:d\vec k\;\psi(\omega,\vec
  k)\:a^\dag\!(\omega,\vec k)|0\>
\labell{psis}\;,
\end{align}
with the same spectrum $\psi(\omega,\vec k)$ as in \eqref{psi}. The
probability of detecting it at time $t$ at transverse position $\vec
r$ on the screen is
\begin{align}
  p({t},\vec r)\propto|\<0|E^+({t},\vec
  r)|\psi_2\>|^2\propto|\tilde\psi({t},\vec r)|^2
\labell{single}\;
\end{align}
In other words, by using the two-photon entangled state $|\psi_2\>$ we
reduced the statistical noise of the time of arrival and of the
transverse position by a half with respect to what one would have
obtained from a single photon $|\psi_1\>$ with identical spectral
function $\psi(\omega,\vec k)$. Clearly, a fair comparison must be
between the two-photon entangled strategy and an unentangled strategy
that uses {\em two} unentangled photons $|\psi_1\>\otimes|\psi_1\>$.
If each of the unentangled photons provide an error equal to the
standard deviation of $|\tilde\psi|^2$, the standard deviation of the
average time of arrival gains a factor of $1/\sqrt{2}$ (because the
variance of the sum is the sum of variances), and similarly for each
of the two components of the average position. Thus, using a
biphoton entangled state $|\psi_2\>$ one obtains a net gain equal to the
square root $\sqrt{2}$ of the number of photons in the resolution
along each of the three spatial directions with respect to a strategy
that employs two unentangled photons $|\psi_1\>$.

It is easy now to extend the above discussion to an arbitrary number $N$
of photons: the joint probability of detecting them at time ${t}_j$ at
transverse position $\vec r_j$ is
\begin{align}
  p(\{{t}_j,\vec r_j\}_{j=1\cdots N})\propto|\<0|\prod_{j}E^+({t}_j,\vec
  r_j)|\psi_N\>|^2\nonumber\\\propto|\tilde\psi(\sum_j{t}_j-Nt_0,\sum_j\vec
  r_j-N\vec r_p)|^2
\labell{cason}\;,
\end{align}
if one uses a far-field $N$-photon entangled state
\begin{align}
  |\psi_N\>\equiv\int d\omega\:d\vec k\;\psi(\omega,\vec
  k)(a^\dag(\omega,\vec k))^N|0\>
  \labell{psin}\;.
\end{align}
Clearly, \eqref{cason} gives a distribution that has a standard
deviation for each position component and for the time of arrival that
is $\sqrt{N}$ times smaller than the standard deviation obtained by
averaging $N$ unentangled photons in the state $|\psi_1\>$, with
arrival probability \eqref{single}.

We now discuss the feasibility of the experiment. For the
state $|\psi_N\>$, the arrival time ${t}_j$ and position $\vec r_j$ of
each photon is completely random. In fact, consider the case
$N=2$: $|\psi_2\>$ can be written also as
\begin{align}
|\psi_2\>=\int dt_1\:d\vec r_1\:dt_2\:d\vec r_2\nonumber
\:\tilde\psi({t}_1+{t}_2,\vec r_1+\vec r_2)\:\\\times
a^\dag\!({t}_1,\vec r_1)
\:a^\dag\!({t}_2,\vec r_2)|0\>
\labell{psift}\;,
\end{align}
where we introduced into \eqref{psi} the operator $a^\dag({t},\vec
r)\propto\int d\omega\:d\vec k\:a^\dag(\omega,\vec k)\:e^{i(\omega
  {t}+\vec k\cdot\vec r)}$ that creates a photon at time $t$ and
transverse position $\vec r$.  Each of the two photons in
\eqref{psift} taken by themselves can arrive at any time and at any
position, since the time and position difference have uniform
probability amplitude. It is only the time and position sums (or
averages) that are peaked.\togli{Here I'm just looking at the input
  state, not at the detected photons!!!! This is not right!!! I should
  redo this calculation by taking the marginal of $p({t}_1,\vec
  r_1;{t}_2,\vec r_2)$ over one of the photons: $p({t}_1,\vec
  r_1)=\int dt_2d\vec r_2p({t}_1,\vec r_1;{t}_2,\vec r_2)$.} Indeed,
the probability \eqref{fin2} depends only on the sums ${t}_1+{t}_2$
and $\vec r_1+\vec r_2$, so that the differences ${t}_1-{t}_2$ and
$\vec r_1-\vec r_2$ must be uniformly distributed.

So, there are two main practical issues with this protocol. On one
hand, it is very demanding to produce the maximally-entangled states
\eqref{psi} and \eqref{psin}. On the other hand, the complete
randomness in arrival times and positions require an infinite
measurement time and transverse screen. Both these problems can be
overcome by reducing the amount of entanglement among photons. This,
of course, will reduce the resolution gain, but it will still allow
for a better-than-classical enhancement. Again, for the sake of
illustration, we will consider the case $N=2$ first, and then extend
to arbitrary $N$.

Consider the partially-entangled two-photon state \begin{align}
|\phi_2\>\equiv\int d\omega\:d\vec k\:d\omega_d\:d\vec
k_d\:\psi(\omega,\vec k)\:\gamma(\omega_d)\:\xi(\vec
k_d)\:\times\nonumber
\\ a^\dag\!(\omega,\vec k)\:a^\dag\!(\omega+\omega_d,\vec k+\vec
k_d)|0\>
\labell{phi2}\;,
\end{align}
where $\omega_d$ and $\vec k_d$ are the frequency difference and
transverse wave vector divergence between the two photons, governed by
the probability amplitudes $\gamma$ and $\xi$ respectively. The state
$|\phi_2\>$ is normalizable and tends to $|\psi_2\>$ in the limit when
$\gamma$ and $\xi$ tend to  delta functions $\gamma\to\delta(\vec
k_p)$, $\xi\to\delta(\omega_d)$. Replacing $|\psi_2\>$
with $|\phi_2\>$ into \eqref{joint}, we find
\begin{eqnarray}
&&p({t}_1,\vec r_1;{t}_2,\vec r_2)\propto
|\tilde\psi({t}_1+{t}_2-2t_0,\vec r_1+\vec r_2-2\vec r_p)|^2\times
\nonumber\\&&\quad|\tilde\gamma({t}_2-{t}_0)\:\tilde\xi(\vec r_2-\vec
r_p)+\tilde\gamma({t}_1-{t}_0)\:\tilde\xi(\vec r_1-\vec r_p)|^2
\labell{phi2joint}\;,
\end{eqnarray}
where $\tilde\gamma$ and $\tilde\xi$ are the Fourier transforms of
$\gamma$ and $\xi$. In the limit in which $\gamma$ and $\xi$ are
deltas, then $\tilde\gamma$ and $\tilde\xi$ are uniform, so the second
line of \eqref{phi2joint} is a constant and we reobtain the maximally
entangled result of \eqref{fin2}. The opposite limit of constant
$\gamma$ and $\xi$ corresponds to the case in which one photon has
spectrum $\psi$ and the other photon has infinite temporal and spatial
bandwidth. In this case, $|\tilde\gamma|^2$ and $|\tilde\xi|^2$ are
deltas and we obtain $p({t}_1,\vec r_1;{t}_2,\vec r_2)\propto
|\tilde\psi({t}_1-{t}_0,\vec r_1-\vec r_0)|^2\delta(\vec r_2-\vec
r_p)\delta({t}_2-{t}_0)+ |\tilde\psi({t}_2-{t}_0,\vec r_2-\vec
r_0)|^2\delta(\vec r_1-\vec r_p)\delta({t}_1-{t}_0)$, which is the
joint distribution one expects when one photon (the one with infinite
bandwidth) determines the position of the object exactly, whereas the
other finds it with probability \eqref{single}. In the intermediate
case \eqref{phi2joint} in which the differences $\omega_d$ and $\vec
k_d$ have finite bandwidth, there are two competing effects: on one
hand the average time $\tfrac{{t}_1+{t}_2}2$ and average position
$\tfrac{\vec r_1+\vec r_2}2$ have a distribution that is wider than
$\tilde\psi$, so these quantities are determined with a lower
resolution than the maximally entangled case. On the other hand, the
marginal distributions of the times ${t}_j$ and positions $\vec r_j$
are not uniform anymore: in the first term of the second line of
\eqref{phi2joint} the distance between ${t}_2$ and ${t}_0$ cannot be
much larger than the standard deviation of $|\tilde\gamma|^2$, and
thus also the distance between ${t}_1$ and ${t}_0$ cannot be too
large, since ${t}_1+{t}_2-2t_0$ has a width governed by
$|\tilde\psi|^2$. Analogously the distance between $\vec r_1$, $\vec
r_2$ and $\vec r_p$ is limited by the standard deviations of
$|\tilde\xi|^2$ and $|\tilde\psi|^2$, and similar considerations apply
to the second term. In essence, each of the photon's time of arrival
${t}_j$ and transverse position $\vec r_j$ is limited (in contrast to
the maximally entangled case), but the spread in their averages is
dominated by the product between $|\tilde\psi|^2$ and
$|\tilde\gamma\tilde\xi|^2$. For such non-maximal entangled states
Eq.(\ref{phi2}), the standard deviation of the average time of arrival
gains a factor of $\lambda$ with $1/2\leqslant \lambda\leqslant 1$,
and similarly for each of the two components of the average position.
When the bandwidth of $\xi$ and $\gamma$ is larger than that of
$\psi$, $\lambda\leqslant1/\sqrt{2}$, it will always achieve a
better-than-classical enhancement both in time and transverse
positions.

The $N$-photon extension for the non-maximally entangled state is now
straightforward: use the state
\begin{widetext}
\begin{align}
|\phi_N\>\equiv\int d\omega\:d\vec k\prod_j{d\omega}_j\:{d\vec
k}_j\:\psi(\omega,\vec
k)\:\gamma({\omega}_1)\cdots\gamma({\omega}_N)\:\xi({\vec
k}_1)\cdots\xi({\vec k}_N)\:
a^\dag\!(\omega,\vec k)\:a^\dag\!(\omega+{\omega}_1,\vec k+{\vec
k}_1)\cdots a^\dag\!(\omega+{\omega}_N,\vec k+{\vec
k}_N)|0\>
\nonumber
\end{align}
to calculate
\begin{align}
  p(\{{t}_j,\vec r_j\}_{j=1\cdots N})\propto|\<0|\prod_{j}E^+({t}_j,\vec
  r_j)|\phi_N\>|^2\propto|\tilde\psi(\sum_j{t}_j-Nt_0,\sum_j\vec
  r_j-N\vec r_p)\sum_j\prod_{n\neq
    j}\tilde\gamma({t}_n-{t}_0)\:\tilde\xi(\vec{r}_n-\vec{r}_p)|^2
\label{phiNjoint}\;,
\end{align}
\end{widetext}
for which considerations analogous to the case $N=2$ seen above apply. In the intermediate case \eqref{phiNjoint}
in which the differences $\{\omega_n\}$ and $\{\vec k_n\}$ have finite
bandwidth, there are two competing effects: on one hand the average
time $\sum_j{t}_j$ and average position $\sum_j\vec
  r_j$ have a distribution that is wider than $\tilde\psi$, so these
quantities are determined with a lower resolution than the maximally
entangled case. In essence, each of the
photon's time of arrival ${t}_j$ and transverse position $\vec r_j$ is
limited, but the spread in their averages is dominated by the product
between $|\tilde\psi|^2$ and
$|\prod_{n}\tilde\gamma_n\tilde\xi_n|^2$. For the  non-maximal
entangled states $|\phi_N\>$, the standard deviation of the average time of arrival gains a factor of $\lambda$ with $1/N\leqslant \lambda\leqslant 1$, and similarly for each of the
two components of the average position. When the bandwidth of $\xi$ and $\gamma$ is larger than that of $\psi$, $\lambda\leqslant1/\sqrt{N}$, it will always achieve a better-than-classical enhancement both in time and transverse positions.

The the ideal state $|\psi_N\>$ and $|\phi_N\>$ for arbitrary $N$ is
actually a state that is positively correlated both in frequency {\em
  and transverse momentum}. For $N=2$, the state $|\psi_2\>$ has been
experimentally realized under a tightly focused pulsed pump based on
type II noncritical phase matching \cite{Liu}. Pulsed pumping can
provide the bandwidth for the frequency correlation
($\omega_1=\omega_2$), and a tightly focused process can modulate the
transverse momentum correlation ($\vec k_1=\vec k_2$). According to
the ideal phase matching relation \cite{Liu}, a maximal
positively-correlated momentum source requires an infinitely long
crystal ($L\rightarrow +\infty$) and extremely narrow beam waist
$w_0\rightarrow 0$, where $w_0$ is waist radius of pump at the
entrance to the crystal. $\theta(z)$ represents the variation of the
pump's phase, which depends on the propagation length and the confocal
length of the pump, where the confocal length of the pump is $b=w^2_0
k_{p}$ and $k_{p}$ is the pump wave vector. We define a focal
parameter $\chi=\frac{L}{b}$. When $\chi\ll 1$, i.e., $L\ll b$, the
pump is considered to be collimated where the effect of phase $\theta$
can be neglected, it leads to generate a state which possesses a
negatively-correlated momentum from spontaneous parametric
down-conversion. While $\chi\gg 1$, i.e., $L\gg b$, the effect of
phase $\theta$ plays an important role, which leads to generate a
state which possesses a positively-correlated momentum from
spontaneous parametric down-conversion. Hence, to obtain
positive-correlation in momentum, the requirement $\chi\gg 1$ should
be satisfied, which can be realized by increasing the length of
crystal $L$ and decreasing the beam waist $w_0$. In realistic
situations, these ideal requirements are not met and one obtains the
partially-entangled states $|\phi_2\>$ and $|\phi_N\>$, which are the ones required for our proposal. One
can generate a relatively high quality non-maximally entangled state
by using centimeter-sized periodically poled materials and feasible
focal parameters of the pump. As illustrated in \cite{Liu}, taking the
SPDC process in a KTiOPO4 (KTP) crystal as the example, a relatively
high quality positively-correlated entangled state can be generated
with the crystal length $L=5cm$ and beam waist $w_0=10\mu m$
($\chi=35.81$). Moreover, optical superlattice technologies seem to
have great potential in generating high quality sources \cite{Yu}.

We now briefly consider the effect of noise. The maximally entangled
protocol is extremely sensitive to noise, as typically happens in
quantum metrology: the loss of a single photon will render all the
other $N-1$ ones completely useless for the estimation, since their
times and positions of arrival are completely random. This is the
typical scenario in quantum metrology in the presence of noise
\cite{rafalguta,davidov}, but many different strategies that reduce
the effect of noise at the cost of a slight decrease in resolution
have been proposed. For example, the non-maximally entangled state
$|\phi_N\>$  is more robust to the loss of photons: the
photons that do arrive, still contain partial information on the
object position. As another example, the strategies proposed in
\cite{qps1} can be adapted to the current case. The main idea is that
one divides the $N$ photons into subsets of $M$ entangled photons and
then entangles these subsets among each other (a nested
strategy). Then if one photon is lost, only the photons of its subset
become useless, while those of the other subsets can still attain a
better-than-classical resolution. Other possible strategies involve
the use of quantum error correcting codes \cite{qec} or the use of ancillary
systems that do not participate to the estimation procedure \cite{dep}.

In conclusion, we have proposed a quantum estimation protocol to
estimate the location of a target in three dimensions with a precision
increase equal to the square root of the number of photons employed,
when compared to the best unentangled strategy using photons with
equal spectral characteristics. In this paper we have focused on
entanglement among photons, but quantum squeezing would allow a
similar enhancement \cite{sq}. As a future application, one might
consider the extension of the protocol to the localization in
four-dimensional spacetime to determine the spatial location and the
time of an event. Unfortunately such extension is nontrivial because
in electromagnetic waves the spatial and temporal degrees of freedom
are connected (they are constrained by being a solution to a wave
equation). So one would need a further, independent, degree of freedom
to use as a clock, in addition to the photon's spatial degrees of
freedom that we used here.

\vskip 1\baselineskip L.M. acknowledges funding from Unipv, ``Blue
sky'' project - grant n.~BSR1718573; C.L.R. acknowledge the funding
from National key research and development program
(No.~2017YFA0305200), the Youth Innovation Promotion Association (CAS)
(No.~ 2015317), the National Natural Science Foundation of China (No.
11605205), the Natural Science Foundation of Chongqing (No.
cstc2015jcyjA00021, cstc2018jcyjAX0656), the Entrepreneurship and
Innovation Support Program for Chongqing Overseas Returnees
(No.cx2017134, No.cx2018040), the fund of CAS Key Laboratory of
Microscale Magnetic Resonance, and the fund of CAS Key Laboratory of
Quantum Information.

\end{document}